# Strain-Engineering Anisotropic Electrical Conductance of Phosphorene and Few-Layer Black Phosphorus

Ruixiang Fei and Li Yang[*]

*Department of Physics, Washington University in St. Louis, St. Louis, MO 63130, USA*

**Abstract:** Newly fabricated monolayer phosphorene and its few-layer structures are expected to be promising for electronic and optical applications because of their finite direct band gaps and sizable but anisotropic electronic mobility. By first-principles simulations, we show that this unique anisotropic conductance can be controlled by using simple strain conditions. With the appropriate biaxial or uniaxial strain, we can rotate the preferred conducting direction by 90 degrees. This will be of useful for exploring quantum Hall effects, and exotic electronic and mechanical applications based on phosphorene.

Strain has been known as an effective mechanism for controlling electronic, transport, and optical properties of semiconductors for decades [1-6]. This tool is particularly useful when engineering one-dimensional (1D) and two-dimensional (2D) crystals because these reduced-dimensional structures can sustain much larger strains than bulk crystals. For example, monolayer graphene and $MoS_2$ have been reported to be strained up to their intrinsic limit (~ 25%) without substantially damaging their crystal structures [7,8], providing a dramatically-wide range for tuning their mechanical and electronic performances. Furthermore, beyond device applications, spatially modulated strain can even mimic gauge fields that are crucial for studying more fundamental phenomena in 2D structures, e.g., realizing ultra-strong magnetic fields and associated zero-field quantum Hall effects as observed in graphene [9].

Recently, a promising 2D semiconductor, phosphorene, was successfully fabricated [10-12]. Unlike the widely studied semimetallic graphene, phosphorene exhibits a finite and direct band gap within an appealing energy range [10, 11, 13]. Particularly notable features of phosphorene include its anisotropic electric conductance and optical responses [11, 13], distinguishing this material from many other isotropic 2D crystals, *i.e.*, graphene and molybdenum and tungsten chalcogenides. In this regard, it will be useful to manipulate these anisotropies. Considering that all electronic and optical anisotropies are essentially decided by geometries of atomistic structures and that strain is the most direct way to change the atomistic and electronic structures [14], it is obvious to attempt to use strain to control the appealing anisotropies of phosphorene.

In this work, through first-principles simulations, we show that strain can serve as an efficient tool for controlling the anisotropic electrical conductance of phosphorene. As shown in Fig. 1 (a), the spatial preference of conductance can be rotated by 90 degrees in plane through either uniaxial (less than 6%) or biaxial strain (less than 4 %), which are well achievable given current experimental capabilities. Moreover, our calculated mobilities of electrons and holes exhibit different responses to the external strain; only the anisotropy of the electron conductance is tuned by the strain. Quantum-mechanics calculations further show that this change of electron conducting anisotropy is a result of a switch in the energy order of the first and second lowest-energy conduction bands that is induced by strain. Our observed strain-engineered anisotropic conductance provides many opportunities for studying novel mechanical-electronic devices and unusual quantum Hall effects related to strongly anisotropic effective masses [15].

The studied structures of monolayer and bilayer phosphorene are presented in Figs. 1 (b), (c) and (d). Each phosphorus atom is covalently bonded with three neighboring phosphorus atoms to form a puckered 2D honeycomb structure. We fully relax these structures according to the force and stress calculated by density functional theory (DFT) with the Perdew, Burke and Ernzerhof (PBE) functional [16]. The band energy is obtained by solving the Kohn-Sam equation under plane-wave basis with normal-conserving pseudopotentials [17]. The plane-wave energy cutoff is set to be 25 Ry and the k-point sampling grid is 14 x 10 x 1, producing converged results. The vacuum spacing between neighboring supercells is set to be 2.5 nm to avoid artificial interactions. Our calculated lattice constants and geometries are similar to previous works on monolayer and few-layer phosphorene [13, 18, 19].

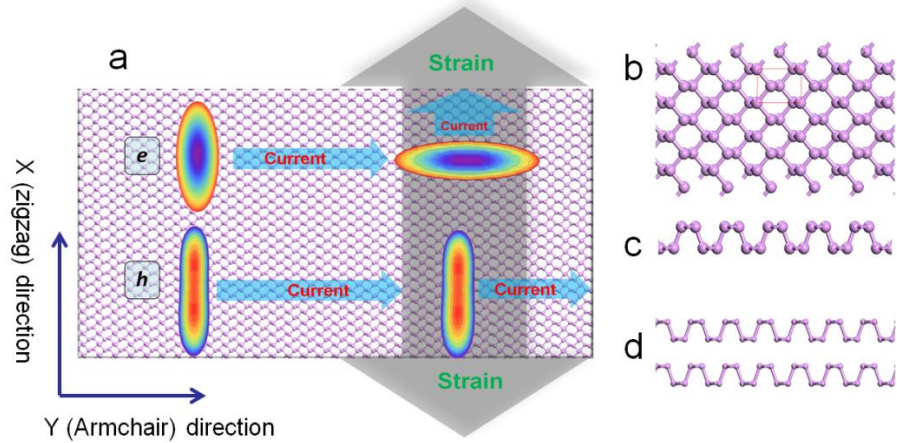

Figure 1 (a) schematic of strain-induced rotation of electrical conductance in monolayer phosphorene. The color contours are those of 2D electron and hole bands. (b) Top view of monolayer phosphorene. (c) and (d) Side views of monolayer and bilayer phosphorene, respectively.

We start from the electronic band structure of intrinsic monolayer phosphorene presented in Fig. 2 (a). In addition to the direct band gap located at the ・ point, an impressive feature of phosphorene is its highly anisotropic band dispersion around the band gap. Both the top of valence bands and the bottom of conduction bands have much more significant dispersions along the ・-Y direction, which is the armchair direction in real space as indicated in Fig. 1 (a); however, these bands are nearly flat along the ・-X (the zigzag) direction. Therefore, the corresponding effective mass of electrons and holes are also highly anisotropic. The anisotropic effective mass of electrons of intrinsic monolayer phosphorene is illustrated in real space in Fig. 3 (a). For different directions, these values can differ by an order of magnitude. This anisotropic effective mass or band dispersion is responsible for the recently-observed anisotropic electric conductance in phosphorene [11].

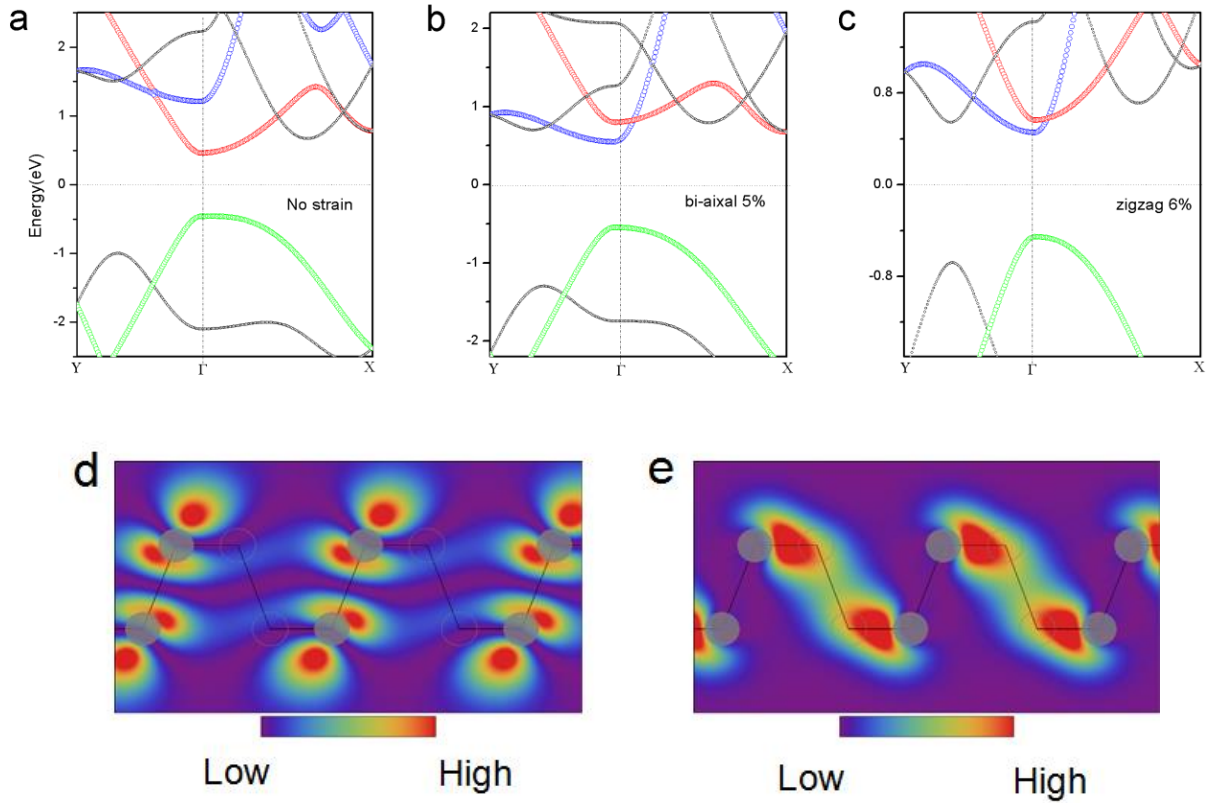

Figure 2 Band structures of: (a) intrinsic monolayer phosphorene, (b) phosphorene with a 5% biaxial strain, and (c) phosphorene with a 6% zigzag uniaxial strain. (d) The side view of the electronic wave function of the lowest-energy conduction band at the Γ point. (e) That of the second-lowest-energy conduction band at the Γ point.

Interestingly, beyond the first conduction band (red-color marked one in Fig. 2 (a)), there is another unique conduction band close to the bottom of the first conduction band, which is marked by blue color in Fig. 2 (a). Excitingly, the dispersion of this blue-colored band is also highly anisotropic, but its spatial isotropy is orthogonal to that of the first conduction band; it is highly dispersive along the ·-X (zigzag) direction while almost flat along the ·-Y (armchair) direction. To understand the anisotropic band dispersion and conductance, we have plotted the real-space wave functions of the first and second conduction bands at the · point in Figs. 2 (d) and (e). It is easy to see the spatial continuity of the wave function along the armchair direction in Fig. 2 (d), which hints a high mobility along this direction. In Fig. 2 (e), the wave function is isolated within each unit cell with a small overlap between each other, indicating low mobility along the armchair direction.

Since the electrical transport behavior is usually decided by the lowest-energy band edge, if we can switch the energy order of the red-colored and blue-colored conduction bands of phosphorene, the anisotropic electrical conductance will change accordingly. Applying external strain proves to be an appealing way to achieve this switch in band order. Let us first focus on the uniform biaxial strain. We do find that this strain lowers the second (blue-colored) conduction band while increases the first (red-colored) conduction band. In Fig. 2 (b), when the applied biaxial strain is 5%, the energy order of these two conduction bands is switched at the ・ point. Correspondingly, the effective mass of free electrons, which is shaped by the bottom of the conduction band, also rotates its spatial isotropy. As concluded in Fig. 3 (b), the spatial anisotropy of free electrons is rotated by exactly 90 degrees; now the electrons have a very small effective mass along the ・-Y (armchair) direction and a large effective mass along the ・-X (zigzag) direction.

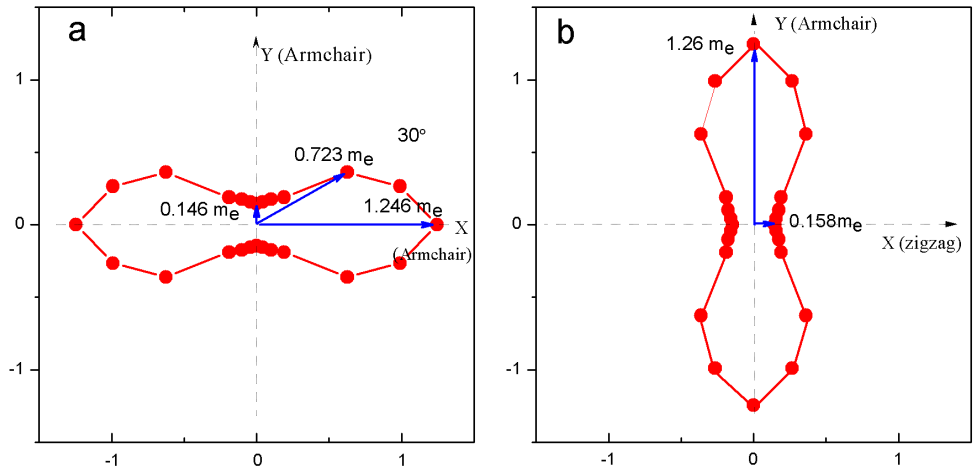

Figure 3.  The electron effective mass according to spatial direction. (a) That of intrinsic phosphorene and (b) that of 5% biaxially strained phosphorene. The length of the blue arrow represents the absolute value of effective mass.

In order to directly connect the anisotropic band dispersion to the electrical conductance, we have further estimated the electrical mobility of electrons along specific directions (zigzag and

armchair directions) according to the formula [20-23]:

$$\mu_{x\_anisotropy} = \frac{eh^3 C_{x\_anisotropy}}{(2\pi)^3 k_B T m_e^* m_d E_{1x}^2} \quad (1)$$

Here $m_e^*$ is the effective mass along the transport direction, and the density-of-state mass $m_d$ is determined by $m_d = \sqrt{m_{ex}^* m_{ey}^*}$. The deformation potential constant $E_{1x} = \Delta E/(\Delta l_x / l_x)$ is obtained by varying the lattice constant along transport direction $\Delta l_x$ ($l_x$ is the lattice constant along the transport direction, here is x direction) and checking the change of band energy under the lattice compression and strain. The elastic module $C_{x\_anisotropy}$ is obtained by $C_{x\_anisotropy}(\Delta l_x / l_x)^2 / 2 = (E - E_0)/S_0$, where $E - E_0$ is obtained by varying the lattice constant by small amount ($\Delta l_x / l_x \sim 0.5\%$) to obtain the change of the total energy, and $S_0$ is the lattice area in xy plane. $T$ is the temperature. The estimation only considers the simplest electron-phonon coupling and gives the upper limit of mobility of realistic cases, in which many other extrinsic factors will further decrease these overestimated mobility values. However, this formula is sophisticated enough to capture the anisotropic conductance of this system.

The calculated electron mobility at temperature $T = 300K$ according to the biaxial strain is shown in Fig. 4 (a). First, for intrinsic phosphorene, the mobility of electron along the zigzag direction is impressive, around 2000 cm$^2$/v·s, which is much larger than those observed in other 2D semiconductors, such as MoS$_2$ [24]. This is also consistent with previous predictions and measurements, which claim that phosphorene may be a promising material for various high-performance devices [23]. Second, for different percentages of biaxial strain, the critical transition between anisotropic mobilities occurs at a strain between 3% and 4%, which is well within current experimental capabilities. After the switch, the mobility along the zigzag direction is around 10 times of magnitude larger than that along the armchair direction, directly showing the rotation of the anisotropic electrical conductance.

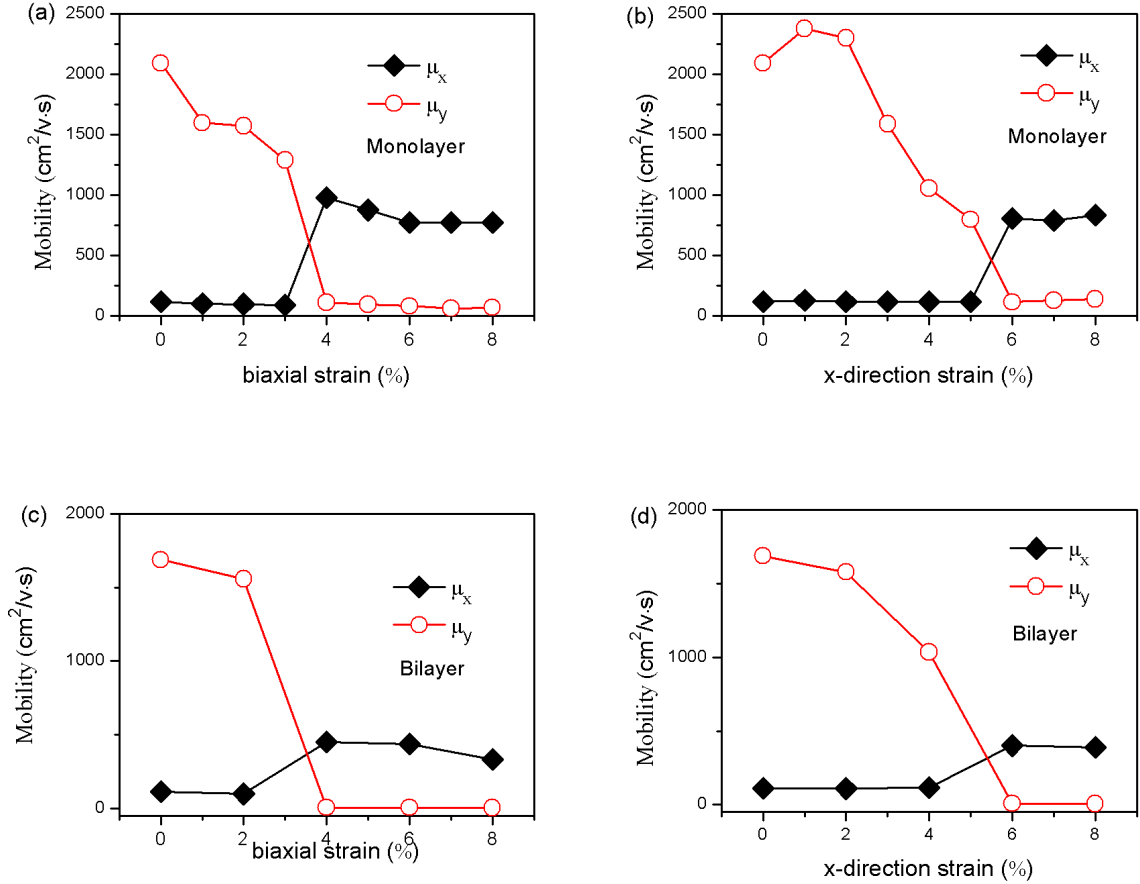

Figure 4. (a) and (b) Electron mobilities of monolayer phosphorene along the zigzag (x) and armchair (y) directions for biaxial strain and uniaxial strain along the zigzag direction, respectively. (c) and (d) The electron mobilities of bilayer phosphorene along the zigzag (x) and armchair (y) directions for biaxial strain and uniaxial strain along the zigzag direction, respectively. The black diamond and red hollow circle indicate the x-direction electron mobility and y-direction electron mobility, respectively.

In addition to biaxial strain, we also observe the similar rotation of the electric conductance for uniaxial strain on phosphorene. As shown in Fig. 2 (b), the energy order of the first two conduction bands can be switched by applying uniaxial strain along the zigzag direction. The critical strain magnitude (between 5% and 6%, reading from Fig. 4 (b)) for rotating the electrical conductance is slightly larger than that of biaxial strain (between 3% and 4%). On the other hand, applying uniaxial strain along the armchair direction does not produce a band switch and thus there is no rotation of the electrical conductance.

Finally, we observe a similar rotation of the anisotropy of electrical conductance for bilayer phosphorene as well. In Figs. 4 (c) and (d), the electron mobility values along the zigzag and armchair directions switch for external uniaxial strains above 5%, or for biaxial strains larger than 3%, which are similar to those in monolayer structures. Therefore, this strain-tunable anisotropy of the electrical conductance can also be realized in multi-layer phosphorene structures.

In conclusion, we have demonstrated a novel mechanism for engineering the unique, anisotropic electrical conductance in monolayer and few-layer phosphorene. By applying appropriate uniaxial or biaxial strains, the anisotropy of the electron effective mass and corresponding mobility direction can be rotated by 90 degrees. First-principles simulations reveal that this is a result of a switch in the energy order of the lowest two conduction bands. This discovery of a method for controlling the exotic anisotropies of phosphorene makes this material even more promising for mechanical and electronic applications.

This work is supported by the National Science Foundation Grant No. DMR-1207141. The computational resources have been provided by the Lonestar of Teragrid at the Texas Advanced Computing Center (TACC). The DFT calculation is performed with the Quantum Espresso [25].

* lyang@physics.wustl.edu

**References**

[1] M. V. Fischetti, Z. Ren, P. M. Solomon, M. Yang, and K. Rim, J. Appl. Phys. 94, 1079 (2003).

[2] Mistry, Kaizad, et al., Electron Devices Meeting, 2007. IEDM 2007. IEEE International. IEEE, 2007.

[3] Jacobsen, Rune S., et al. "Strained silicon as a new electro-optic material." Nature 441.7090 (2006): 199-202.

[4] Haeni, J. H., et al. "Room-temperature ferroelectricity in strained SrTiO3." Nature 430.7001 (2004): 758-761.

[5] Falvo, M. R., et al. "Bending and buckling of carbon nanotubes under large strain." Nature 389.6651 (1997): 582-584.

[6] Lai, Keji, et al. "Mesoscopic percolating resistance network in a strained manganite thin film." Science 329.5988 (2010): 190-193.

[7] C. Lee, X. Wei, J. W. Kysar, J. Hone, Science 321, 385 (2008).

[8] S. Bertolazzi, J. Brivio, A. Kis, ACS Nano 5, 9703 (2011).

[9] F. Guinea, M. I. Katsnelson, A. K. Geim，Nature Physics, 6, 30 (2010).

[10] Likai Li, Yijun Yu, Guo Jun Ye, Qingqin Ge, Xuedong Ou, Hua Wu, Donglai Feng, Xian Hui Chen, and Yuanbo Zhang, arXiv:1401.4117 (2014).

[11] Han Liu, Adam T. Neal, Zhen Zhu, David Tomanek, and Peide D. Ye, arXiv:1401.4133 (2014).

[12] Fengnian Xia, Han Wang, Yichen Jia, arXiv:1402.0270 (2014).

[13] Vy Tran, Ryan Soklaski, Yufeng Liang, and Li Yang, arXiv:1402.4192 (2014).

[14] A. S. Rodin, A. Carvalho, A. H. Castro Neto, arXiv:1401.1801 (2014).

[15] Sunanda P. Koduvayur, etc., Phys. Rev. Lett. 106, 016804 (2011).

[16] John P. Perdew, Kieron Burke, and Matthias Ernzerhof, Phys. Rev. Lett. 77, 3865 (1996).

[17] N. Troullier and J. L. Martins, Phys. Rev. B 43, 1993 (1991).

[18] Allan Brown and Stig Rundqvist, Acta Cryst 19, 684 (1965).

[19] Y. Takao, H. Asahina, and A. Morita, Journal of Phys. Soc. of Jap. 50, 3362 (1981).

[20] S. Bruzzone and G. Fiori, Applied Physics Letters 99, 2108 (2011).

[21] S.-i. Takagi, A. Toriumi, M. Iwase, and H. Tango, IEEE Transactions on 41, 2357-2362 (1994).

[22] G. Fiori, and G. Iannaccone, Proceedings of the IEEE 101, 1653 (2013).

[23] Jingsi Qiao, Xianghua Kong, Zhi-Xin Hu, Feng Yang, and Wei Ji, arXiv:1401.5045

(2014).

[24] B. Radisavljevic, A. Radenovic, J. Brivio, V. Giacometti, and A. Kis, Nature nanotech. **6**, 147 (2011).

[25] P. Giannozzi *et al.*, J. Phys.: Condens. Matter **21**, 395502 (2009).